\documentclass[twocolumn,amsmath,amssymb,twoside,preprintnumbers,floatfix,showpacs,pre,superscriptaddress]{revtex4}

\usepackage[latin1]{inputenc}
\usepackage{vmargin}
\usepackage{graphicx}
\usepackage[dvips]{epsfig}

\usepackage{amsmath}
\usepackage{amssymb}
\usepackage{dcolumn}
\usepackage{bm}

\graphicspath{{}{images/}}

\newcommand{\eq}{Eq.\,}
\newcommand{\eqns}{Eqns.\,}

\newcommand{\fig}{Fig.\,}
 
\newcommand{\sect}{Sec.\,}

\def\epsilon{\varepsilon}
\def\theta{\vartheta}
\def\Ref#1{Ref.\,\cite{#1}}
\def\rho{\varrho}

\begin{document}
\title{Shear Viscosity of Clay-like Colloids in Computer Simulations and Experiments}

\author{Martin Hecht}
\affiliation{Institute for Computational Physics,
Pfaffenwaldring 27,
70569 Stuttgart,
Germany}
\author{Jens Harting}
\affiliation{Institute for Computational Physics,
Pfaffenwaldring 27,
70569 Stuttgart,
Germany}
\author{Markus Bier}
\affiliation{
   Max-Planck-Institut f\"ur Metallforschung,  
   Heisenbergstra\ss e 3, 
   70569 Stuttgart, 
   Germany \\
   and 
   Institut f\"ur Theoretische und Angewandte Physik, 
   Universit\"at Stuttgart, 
   Pfaffenwaldring 57, 
   70569 Stuttgart, 
   Germany
}

\author{J\"org Reinshagen}
\affiliation{
Institute of Ceramics in Mechanical Engineering, 
University of Karlsruhe (TH), Karlsruhe, 76131 Germany
}

\author{Hans J. Herrmann}
\affiliation{Institut f\"{u}r Baustoffe, Schafmattstr. 6, 
ETH Z\"{u}rich, CH-8093 Z\"{u}rich, Switzerland}
\affiliation{Departamento de F\'{i}sica, Universidade Federal do Cear\'{a}
Campus do Pici, 
60451-970 Fortaleza CE, Brazil}

\date{\today}

\begin{abstract}
\noindent
\textbf{Abstract.}

Dense suspensions of small strongly interacting particles are complex
systems, which are rarely understood on the microscopic level.
We investigate properties of dense
suspensions and sediments of small spherical Al$_2$O$_3$ particles in a shear
cell by means of a combined Molecular Dynamics (MD) and Stochastic
Rotation Dynamics (SRD) simulation. We study structuring 
effects and the dependence of the suspension's viscosity on the shear rate and 
shear thinning for systems of varying salt concentration and $p$H value. To show
the agreement of our results to experimental data, the relation between bulk
$p$H value and surface charge of spherical colloidal particles is modeled by
Debye-H\"{u}ckel theory in conjunction with a $2pK$ charge regulation model.
\end{abstract}



\maketitle

\begin{section}{Introduction}
We simulate colloids of silt particles, for which in many cases the attractive Van-der-Waals 
forces are relevant. These colloids are sometimes called ``peloids'' (Greek: clay-like). 
In contrast to clays consisting of thin platelets\cite{vanOlphen} our particles 
are in first approximation spherical particles. For real clays the particle
shape and their orientation is of relevance\cite{DiMasi, Agra04, Lekkerkerker00, Dijkstra95, Dijkstra96}. For silt particles on the other hand the description 
is less complex. However, due to the particle size of micro meters and below,
the interplay of diffusion, electrostatic repulsion,
van der Waals attraction, and hydrodynamics still renders the suspension a 
very complex system. Colloid science tries to investigate the properties 
of such suspensions and there is a vast amount of literature on this
subject\cite{Mahanty76,Lagaly97,Shaw92,Morrison02,Hunter01}. Colloids in general
have various applications ranging from food industry over paintings and
cosmetic products to applications in photographic processes. Particles with
well defined properties can be used to investigate general properties of 
soft condensed matter like gelation or crystallization on a larger length 
scale than on the atomic level. Especially attractive interactions (depletion forces
as well as van der Waals attraction) have drawn the attention in the recent 
years\cite{Trappe01, Sator04, Voivod04, Cates04, Mallamace06}.
In soil mechanics real samples, e.g., of 
sediments, can be less characterized and therefore it is more difficult to
gain a microscopic picture from which general properties can be derived.
Therefore we have chosen a synthetic Al$_2$O$_3$ powder suspended in water
as a model system for silt. The particle diameter is $0.37\,\mu$m. 

Al$_2$O$_3$ is not only a cheap testing material for investigations related to soil 
mechanics, but it is also an important material for ceramics. In process engineering 
one of the basic questions is, how to obtain components of a predefined shape. Wet 
processing of suspensions, followed by a sinter process is a common practice 
here.\cite{Reinshagen06}
Nevertheless, to optimize the production process and to improve the homogeneity
and strength of the fabricated workpiece one has to understand the complex rheological
behavior of the suspension and its relation to the microscopic structure.
This knowledge in turn can be applied to soil mechanics. Shear thinning as 
observed in our simulations and experiments is an important mechanism 
for the dynamics of landslides making them more dangerous. 

In this paper we present our simulation results of sheared suspensions of Al$_2$O$_3$ 
particles. The overall behavior is strongly determined by the effective interaction 
potential between the particles in the suspension. The potentials can be related to 
experimental conditions within Debye-H\"uckel theory, and thus we can compare our 
simulation results to experimental data. 
In contrast to our approach of a direct comparison to experimental data, in the 
literature simulation results are often compared to analytical calculations.

Many different simulation methods have been developed and applied to colloidal 
suspensions: Stokesian dynamics (SD)\cite{Brady88, Brady93, Brady96}, 
accelerated Stokesian dynamics (ASD)\cite{Brady01, Brady04}, pair drag 
simulations\cite{Silbert97a}, Brownian Dynamics (BD)\cite{Huetter99, Huetter00}, 
Lattice Boltzmann (LB)\cite{Ladd94, Ladd94b, Ladd01, Harting04}, 
and Stochastic Rotation Dynamics (SRD)\cite{Inoue02, Padding04, Hecht05}.
Due to the complex nature of the problem, all simulation methods have to simplify 
in some point. Either Brownian motion is neglected or hydrodynamic interactions
are included on a very simplified level. In many cases simulations are done 
without a \emph{quantitative} comparison to experiments. 
In the present paper we combine Molecular Dynamics (MD) to simulate the colloidal
particles, SRD for the description of the fluid, and
a charge regulation model which provides us with realistic parameters for the 
Derjaguin Landau Vervey Overbeek (DLVO)
potentials\cite{DLVO, DLVO2} in the MD simulation. We include long range hydrodynamic
interactions on a coarse grained level in the SRD part and we only include DLVO 
pair-potentials in the MD part. No electrostatic many body interactions or 
electrodynamic interactions 
are considered and modifications of the pair potentials due to locally increased 
colloid concentrations are neglected, too. However, many numerical investigations
are based on much simpler models than ours. To our opinion our model covers the 
main properties quite well.

Our paper is organized as follows: first we shortly describe our MD implementation 
followed by a short sketch of the SRD simulation method, and a description of how we 
have implemented 
our shear cell. The simulation method is described in detail in \Ref{Hecht05}.
Then we describe the so called $2pK$ charge regulation model which relates
our simulation parameters with the $p$H-value and the ionic strength $I$ adjusted in 
the experiment. A short description of the simulation setup and of the experiments carried 
out follows. After that we present our simulation results and compare them
to the experimental data. Finally a summary is given.
\end{section}

\begin{section}{Molecular Dynamics}
We study colloidal particles, composing the solid fraction, suspended in a fluid
solvent. The colloidal particles are simulated with Molecular Dynamics (MD), whereas
the solvent is modeled with Stochastic Rotation Dynamics (SRD) as described in 
\sect\ref{sec_srd}.
\\
In the MD part of our simulation we include effective electrostatic interactions and
van der Waals attraction, a lubrication force and Hertzian contact forces.
The electrostatic and van der Waals potential are usually referred to as DLVO 
potentials\cite{Shaw92,Russel95,Lagaly97,Huetter00,Lewis00,Morrison02}, which capture the 
static properties of colloidal particles in aqueous suspensions. The first component
is the screened Coulomb term
\begin{equation}
 \begin{array}{rcl}
  V_{\mathrm{Coul}} &=& 
  
  \pi \epsilon_r \epsilon_0
  \left[ \frac{2+\kappa d}{1+\kappa d}\cdot\frac{4 k_{\mathrm{B}} T}{z e}
         \tanh\left( \frac{z e \zeta}{4 k_{\mathrm{B}} T} \right)
  \right]^2 \\
  && \times \frac{d^2}{r} \exp( - \kappa [r - d]),
 \end{array}
 \label{eq_VCoul}
\end{equation}
where $d$ denotes the particle diameter and $r$ is the distance between the 
particle centers. $e$ is the elementary charge, $T$ the temperature, 
$k_{\mathrm{B}}$ the Boltzmann constant, and $z$ is the valency of the ions of 
added salt. Within DLVO theory one assumes linear screening, mainly by one 
species of ions with valency $\pm z$ (e.g. $z=+1$ for NH$_4^+$). The first fraction
in \eq\ref{eq_VCoul} is a correction to the original DLVO potential, which takes
the surface curvature into account and is valid for spherical particles\cite{Trizac02}.

The effective surface potential $\zeta$ is the electrostatic potential 
at the border between the diffuse layer and the compact layer, it may therefore be 
identified with \emph{the} $\zeta$-potential. It includes the effect of the bare charge 
of the colloidal particle itself, as well as the charge of the ions in the Stern layer,
where the ions are bound permanently to the colloidal particle. In other words, 
DLVO theory uses a renormalized surface charge, which we determine by the model 
described in \sect\ref{sec_model}. 

$\kappa$ is the inverse Debye length defined by $\kappa^2 = 8\pi\ell_BI$, with the 
ionic strength $I$.  
The Bjerrum length $\ell_B := \frac{\beta e^2}{4\pi\epsilon_0\epsilon_r}$ measures
the distance at which the electrostatic interaction of two elementary charges 
amounts $\beta^{-1} = k_{\mathrm{B}}T$.
$\epsilon_0$ is the permittivity of the vacuum, $\epsilon_r$ the relative dielectric
constant of the solvent (we use 81 for water, i.e., $\ell_B = 7\,\mathrm{\AA}$ for room 
temperature).

The Coulomb term of the DLVO potential competes with the attractive van der Waals term
\begin{equation}
 \begin{array}{rl}
  V_{\mathrm{VdW}} = - \frac{A_{\mathrm{H}}}{12} &
     \left[ \frac{d^2}{r^2 - d^2} + \frac{d^2}{r^2} \right. \\
            & \;\left. + 2 \ln\left(\frac{r^2 - d^2}{r^2}\right) \right].
  \end{array}
\end{equation}
$A_{\mathrm{H}}=4.76\cdot 10^{-20}\,\mathrm{J}$ is the Hamaker constant\cite{Huetter99} 
which involves the polarizability of the particles. The singularity of $V_{\mathrm{VdW}}$ 
for touching particles is removed and the primary minimum is modeled by a parabola as 
described in \Ref{Hecht05}.

Long range hydrodynamic interactions are taken into account in the simulation for
the fluid as described below. This can only reproduce interactions correctly down
to a certain length scale. On shorter distances, a lubrication force has to be
introduced explicitly in the MD simulation.
The most dominant mode, the so-called squeezing mode, is an additional force
\begin{equation}
  \label{eq_FLub}
  \mathbf{F}_{\mathrm{lub}} = -(\mathbf{v}_{\mathrm{rel}},\mathbf{\hat{r}})\mathbf{\hat{r}}
    \frac{6 \pi \eta}{r - d}\left(\frac{R}{2}\right)^2
\end{equation}
between two particles with radius $R$ and relative velocity $\mathbf{v}_{\mathrm{rel}}$.
$\eta$ is the dynamic viscosity of the fluid. In contrast to the DLVO potentials
the lubrication force is a dissipative force. When two particles approach each
other very closely, this force becomes very large. To ensure numerical stability
of the simulation, one has to limit $\mathbf{F}_{\mathrm{lub}}$. 
We chose a maximum force at a certain gap width $r_{\mathrm{sc}}$ and shift 
the force so that the maximum force cannot be exceeded: Instead of calculating 
$\mathbf{F}_{\mathrm{lub}}(r)$ we take the value for 
$\mathbf{F}_{\mathrm{lub}}(r+r_{\mathrm{sc}})$. In addition, since the force decays for 
large particle distances, we can introduce a large cutoff radius $r_{\mathrm{lc}}$ for which
we assume $\mathbf{F}_{\mathrm{lub}}(r) \equiv 0$ if $r - d > r_{\mathrm{lc}}$.
As the intention of $\mathbf{F}_{\mathrm{lub}}$ is to correct the 
finite resolution of the fluid simulation, $r_{\mathrm{sc}}$ and $r_{\mathrm{lc}}$
have to be adjusted in a way that the dynamic properties, i.e., the viscosity
of a simulated particle suspension with weak DLVO interactions fits the measurements.
It turns out that $r_{\mathrm{sc}} = \frac{d}{40}$ and $r_{\mathrm{lc}} = \frac{3}{2} d$
work best. Our approach for $\mathbf{F}_{\mathrm{lub}}$ is similar to the one often
used in lattice Boltzmann simulations\cite{Ladd01}. In contrast to Ladd\cite{Ladd01} we 
have chosen to use two cut-off radii to be able to treat small and large gaps separately.
There are different approaches, e.g., for Stokesian dynamics\cite{Brady88}, where the 
force field is expanded to a multi pole series and the far field part is subtracted 
afterwards.

Finally we use a Hertz force described by the potential
\begin{equation}
  V_{\mathrm{Hertz}} = K (d-r)^{5/2}  \quad  \mathrm{if}  \quad r<d,
\end{equation}
where $K$ is the constant which describes the elasticity of the particles in the
simulation. The Hertz force avoids that the particles penetrate each other.
It also contains a damping term in normal direction,
\begin{equation}
  \mathbf{F}_{\mathrm{Damp}} =  -(\mathbf{v}_{\mathrm{rel}},\mathbf{\hat{r}})\mathbf{\hat{r}}
    \beta_D  \sqrt{d - r},
\end{equation}
with a damping constant $\beta_D$. 

Since in this work no stress perpendicular to the 
shear direction is applied, the tangential forces at the particle surface are not of essential 
importance. To verify this, we have increased the spacial resolution of the 
fluid simulation, included tangential forces on the particles and allowed particle
rotations. Even though the computational effort was considerably larger and one 
could expect that more effects on the length scale below the particle diameter
could be covered. However one could observe only a change of some percent in the viscosity and in 
the velocity profile. Due to the DLVO potential and the lubrication force the particles very 
rarely get into contact as long as no confining stress is applied. 
The only case, in which particles really touch each other, would be if the $\zeta$-potential
is close to zero at a certain $p$H value. This $p$H value is called ``isoelectric point''. 
It depends on the material of the suspended particles and on the solvent. For our system
it is at $p\mathrm{H}=8.7$\cite{Lewis00}. In experiments close to the isoelectric point a 
solid fraction immediately flocculates out and sediments. In the simulation one ends up
with only one big cluster  in the simulation volume, which corresponds to a part of a 
floc seen in the experiment.

For this study we do not apply tangential forces and thus, having only 
central forces, we could neglect rotation of the particles. This reduces the computational
effort substantially. 
\end{section}

\begin{section}{Stochastic Rotation Dynamics (SRD): Simulation of the Fluid}
\label{sec_srd}
The stochastic rotation dynamics method (SRD) was first introduced by Malevanets and Kapral \cite{Malev99, Malev00}. The method is also known as ``real-coded lattice gas'' 
\cite{Inoue02} or as ``multi-particle-collision dynamics'' (MPCD) and has been successfully
applied to simulate many important systems like complex fluids containing polymers 
\cite{Gompper04, yeomans-2004-ali}, vesicles in flow\cite{Gompper04c}, dynamics of chemical reactions\cite{Kapral04}. The method is a promising tool for a coarse-grained description of
a fluctuating solvent, e.g. in \Ref{Lamura01} results of simulations of a flow around a cylinder 
are presented, or in \Ref{Padding04} sedimentation of a particle suspension is studied.

The method is based on so-called fluid particles with continuous positions and velocities.
Each time step is composed of two simple steps: One streaming step and one interaction step.
In the streaming step the positions of the fluid particles are updated as in the
Euler integration scheme known from Molecular Dynamics simulations, 
\begin{equation}
\label{eq_move}
{\bf r}_i(t+\tau)={\bf r}_i(t)+\tau\;{\bf v}_i(t),
\end{equation}
where ${\bf r}_i(t)$ denotes the position of the particle $i$ at time $t$, 
${\bf v}_i(t)$ its velocity at time $t$ and $\tau$ is the time step used for
the SRD simulation. 
After updating the positions of all fluid particles they interact collectively in 
an interaction step which is constructed to preserve momentum, energy and particle number.
The fluid particles are sorted into cubic cells of a regular lattice and only the  
particles within the same cell are involved in the interaction step. First, their
mean velocity ${\bf u}_j(t')=\frac{1}{N_j(t')}\sum^{N_j(t')}_{i=1} {\bf v}_i(t)$
is calculated, where ${\bf u}_j(t')$ denotes the mean velocity of cell $j$ containing
$N_j(t')$ fluid particles at time $t'=t+\tau$. Then, the velocities of each fluid particle in 
cell $j$ are updated as:
\begin{equation}
\label{eq_rotate}
{\bf v}_i(t+\tau) = {\bf u}_j(t')+{\bf \Omega}_j(t') \cdot [{\bf v}_i(t)-{\bf u}_j(t')].
\end{equation}
${\bf \Omega}_j(t')$ is a rotation matrix, which is independently chosen randomly
for each time step and each cell. We use rotations about one of the coordinate axes by 
an angle $\pm\alpha$, with $\alpha$ fixed\cite{Ihle03c}.
The coordinate axis as well as the sign of the rotation are chosen at random, 
 resulting in 6 possible rotation matrices. The mean velocity ${\bf u}_j(t)$ in 
the cell $j$ can be seen as streaming velocity of the fluid at the position of the cell 
$j$ at the time $t$, whereas the difference $[{\bf v}_i(t)-{\bf u}_j(t')]$ entering 
the interaction step can be interpreted as a contribution to the thermal 
fluctuations. Thus, to calculate the local temperature in the cell under 
consideration one has to sum over the squares of this expression.

The method just described is able to reproduce hydrodynamics and thermal fluctuations.
To couple the colloidal particles to the streaming field of the solvent, we use 
``Coupling II'' of \Ref{Hecht05}: we modify the rotation step of the original SRD algorithm 
slightly. The colloidal particles are sorted into the SRD cells as well and their velocity 
enters into the calculation of the mean velocity ${\bf u}_j(t)$ in cell $j$. Since the mass
of the fluid particles is much smaller (in our case it is 250 times smaller) 
than the mass of the colloidal particles, we have to use the mass of each 
particle---colloidal or fluid particle---as a weight factor when calculating the mean 
velocity 
\begin{eqnarray}
\label{eq_rotateMD}
{\bf u}_j(t')&=& \frac{1}{M_j(t')}\sum\limits^{N_j(t')}_{i=1}{\bf v}_i(t) m_i,
\\
\label{eq_rotateMD2}
\mathrm{with}\qquad M_j(t')&=&\sum^{N_j(t')}_{i=1}m_i,
\end{eqnarray}
where we sum over all colloidal and fluid particles in the cell, so that $N_j(t')$ is 
the total number of both particles together. $m_i$ is the mass of the
particle with index $i$ and therefore $M_j(t')$ gives the total mass contained 
in cell $j$ at the time $t'=t+\tau$. The update rule for the particle velocities 
${\bf v}_i(t)$ and positions ${\bf r}_i(t+\tau)$, which we apply, is summarized
in \eqns(\ref{eq_move})--(\ref{eq_rotateMD2}). This method to couple some embedded material
to the SRD simulation is described for different applications in the 
literature\cite{Falck04, Gompper04b}. \\
This coupling method does not enforce no-slip boundaries on the particle surface, as 
in the method suggested by Inoue et al.\cite{Inoue02}. Moreover, as very recently discussed 
by Padding and Louis\cite{Padding06}, purely radial interactions effectively introduce
slip boundary conditions. Considering the drag coefficient, a pre-factor changes
and this could be corrected by assuming a different hydrodynamic radius. We have checked 
the influence of hydrodynamic interactions by removing the fluid completely and by varying the
resolution of the SRD simulation. Also two different coupling methods as described 
in \Ref{Hecht05} have been applied. Without fluid, the achieved shear rate as well as the
viscosity differed strongly, whereas the difference between the two coupling methods 
was in the order of some per cent only. Therefore, we need hydrodynamics to some extend, 
but we have chosen the coupling method with less computational effort. 
Very recently, Yamamoto et al. have shown that for colloidal gelation hydrodynamic 
interactions are of minor importance\cite{Yamamoto06} for 3D systems, but in contrast to our 
work, they focus on the static properties of a colloidal system quenched to zero temperature.

In \Ref{Hecht05} we have described a simple method to introduce shear at the fluid
boundary by adding a velocity offset to all fluid particles reflected at the shear plane. 
From a constant velocity offset $\Delta {\bf v}$ one can calculate the mean 
shear force
\begin{equation}
  \label{eq_shearforce}
  {\bf F}_{\mathrm S}=\left\langle\sum^{L}_{i=1}m_i \frac{\Delta {\bf v}_i(t)}{\tau}\right\rangle,
\end{equation}
where $L$ denotes the average number of fluid particles crossing through the shear plane 
in one time step and $\left\langle\mathrm{\dots}\right\rangle$ stands for a time average. 
$L$ can be expressed by the mean free path and the number density of 
fluid particles. This would be a force driven shear, where one has
only indirect control on the shear rate $\dot\gamma$ or the shear velocity ${\bf v}_S$
respectively. Therefore, we modify the mean velocity ${\bf u}_j(t')$ 
in the cells close to the shear plane by changing the velocity of each fluid particle
as well as the velocity of the colloidal particles contained in that specific cell
by the difference ${\bf v}_S - {\bf u}_j(t')$. By construction the mean 
velocity in these cells is equal to the shear velocity ${\bf v}_S$ after that step.
At the wall itself we implement full slip boundary conditions for the fluid and for 
the colloidal particles. The boundary in the direction of the shear profile 
(direction of the velocity gradient) is chosen to be non-periodic. By doing so, we can 
also observe phenomena like wall-slip, non-linear velocity profiles or density profiles 
in our shear cell (see \sect\ref{sec_viscosity}). In the case of a non-linear velocity profile 
the viscosity is not well defined. We extract the central region of the profile where it is in 
first approximation linear and estimate there an averaged viscosity. This is the ratio
of the velocity gradient and the shear force which can be calculated in analogy to
\eq(\ref{eq_shearforce}) by carrying out the sum over all velocity changes made.
The region where we estimate the velocity gradient is half the system size. 

We have tested a number of boundary conditions and different ways to impose shear, but 
the method just described turned out to work best. No-slip boundaries at a top and
bottom plane seemed to work for high volume fractions and unless the potentials get 
attractive. As soon as (only slight!) cluster formation sets in the particles concentrate 
in the center of the system and lose contact to the sheared walls. Shearing only the fluid
and not the colloidal particles works always, but the resulting viscosity is much too
small. In fact, what one measures is the flow of the fluid streaming around the particles 
like a flow through a porous medium. The next point is how to determine the shear force
and the velocity gradient we need for the calculation of the shear viscosity. 
The force is always related to any velocity changes made in the system and its calculation 
is straightforward in most cases.
\\
The imposed velocity difference divided by the system size perpendicular to the shear 
plane would give an averaged gradient. For clustered systems not even the shape of the 
shear viscosity against shear rate was comparable to the measurements.
If the velocity gradient changes within the system (compare \fig\ref{fig_VxProfiles}), 
we have to take care that we measure the viscosity in the bulk, i.e., that we take 
the velocity gradient there. At least if the particles are not too strongly clustered,
the slope of the plateau in the center of the system can be taken as a ``good'' velocity
gradient. We use this velocity gradient as achieved shear rate as mentioned 
above. With this scheme for strongly attractive forces the obtained viscosity 
$\eta(\dot\gamma)$ for the simulation stays in the vicinity of the measured
curve, whereas for other methods we tried out the points of the simulation usually ended 
up far off the measured curve.
\\
Fully periodic boundary conditions for sheared systems, known as Lees Edwards 
conditions would be a good choice for stable suspensions. As soon as clusters are formed, 
the velocity profile becomes non linear, as discussed above. But, additionally the 
location of the cluster, i.e., the position of the plateau, is not fixed anymore to
the center of the system, which makes it more difficult to extract the correct velocity
gradient. In addition, the shear force would be determined from the velocity changes
of the particles passing around the periodic boundaries. If the cluster by chance 
stays in the center of the system, again only the fluid would be sheared and only 
indirectly, transmitted by the fluid, the force would be exerted on the particles, 
as if with closed boundaries only the walls would move and no sheared regions close to 
the wall were implemented. Together with the periodic boundaries this would lead to large 
fluctuations of the shear force, caused by the 
present position of the cluster. Furthermore, the boundary conditions would have to
be consistent for the MD and for the SRD simulation. For the MD part it is important
that the position, where a particle re-enters the system after passing around the 
periodic boundary, is shifted by $2t{\bf v}_S$ with $t$ being the continuously increasing
simulation time and ${\bf v}_S$ the shear velocity. Additionally this shift
has to be wrapped around the periodic boundaries in shear direction. If we do the 
same with the fluid particles the shift could be any value, not necessarily an
integer multiple of the fluid box size. What we want to point out, without any further 
restrictions, the grid in the SRD rotation step would not anymore be regular in this 
plane, which in addition is the plane where one measures the shear force. To overcome 
this problem, one can restrict the shear rate to values determined by the SRD grid 
size and the SRD time step, but the other difficulties mentioned before remain.

\end{section}

\begin{section}{the charge regulation model}
\label{sec_model}
To determine the effective surface potential which enters the DLVO potential,
we use the model described in the following. In reality, the surface charge is
achieved by adsorption and desorption of charge determining ions leading to an 
electrostatic potential difference between surface and bulk which in turn influences 
ion adsorption. A full description of this regulation of surface charges 
requires two parts: the first part describes the relation between surface 
charge density and surface potential due to the electrolytic environment, 
whereas the second part quantifies the ion adsorption depending on the 
surface concentration of charge determining ions.

Concerning the first part, a relation between the surface charge density $\sigma$ 
and the surface potential $\zeta$ of a charged spherical colloidal particle of 
radius $R$ immersed in an electrolytic environment of relative dielectric constant 
$\epsilon_r$ and ionic strength $I$ is given within Debye-H\"{u}ckel theory 
\cite{Deby23,McQu00} by
\begin{equation}
  \zeta = \frac{R\sigma}{\epsilon_0\epsilon_r(1 + \kappa R)}.
  \label{eq:sigmazeta1}
\end{equation}
As mentioned above, we consider the Stern layer as a part of the surface charge 
and thus, we can identify the effective surface potential in DLVO theory with \emph{the}
$\zeta$-potential and we can thus skip a discussion of bare charge versus effective 
charge\cite{Alexander84, Aubouy03, Trizac03}.

In the second part of our model, the adsorption of charge determining ions on the 
surface of the colloidal particle is described by assuming that the only 
mechanism of adsorption is that of protons ($\mathrm{H}^+$) on surface 
sites ($-\mathrm{S}$). It turned out that this assumption leads to reasonable 
results for surfaces made of $\mathrm{Al_2O_3}$. Adsorption is described by the 
two chemical reactions \cite{Chan75}
\begin{eqnarray}
   -\mathrm{S}^- + \mathrm{H}^+ & \rightleftharpoons & -\mathrm{SH} ,   \\
   -\mathrm{SH}  + \mathrm{H}^+ & \rightleftharpoons & -\mathrm{SH_2^+} ,  
\end{eqnarray}
with the two reaction constants
\begin{eqnarray}
   K_1 & := & \frac{[-\mathrm{S}^-][\mathrm{H}^+]\exp(-\beta e \zeta)}{[-\mathrm{SH}]} ,  \\
   K_2 & := & \frac{[-\mathrm{SH}] [\mathrm{H}^+]\exp(-\beta e \zeta)}{[-\mathrm{SH_2^+}]}. 
\end{eqnarray}

In terms of the surface site concentrations, the total number of surface sites per area and the 
surface charge density are given by 
$N_S = [-\mathrm{S^-}] + [-\mathrm{SH}] + [-\mathrm{SH_2^+}]$ and 
$\sigma = -e[-\mathrm{S^-}] + e[-\mathrm{SH_2^+}]$, respectively. 
Defining $pK_1 := -\log_{10}(K_1)$ and $pK_2 := -\log_{10}(K_2)$ yields the point of
zero charge $p\mathrm{H}_z$, i.e., the $p\mathrm{H}$ value of vanishing surface charge, as
$p\mathrm{H}_z = \frac{1}{2}(pK_1 + pK_2)$. The surface site density $N_S$ and the difference 
$\Delta pK := pK_1-pK_2$ are treated as adjustable parameters.

The above equations lead to the relation
\begin{equation}
   \frac{\sigma}{eN_S} = 
   \frac{\delta\sinh(\psi_N - \beta e \zeta)}
   {1 +  \delta\cosh(\psi_N - \beta e \zeta)}
   \label{eq:sigmazeta2}
\end{equation}
with the Nernst potential $\psi_N := \ln(10)(p\mathrm{H}_z - p\mathrm{H})$ and 
$\delta := 2\cdot 10^{-\frac{\Delta pK}{2}}$.

Equations (\ref{eq:sigmazeta1}) and (\ref{eq:sigmazeta2}) can be solved self-consistently 
for $\zeta$ as a function of $p\mathrm{H}$. For our system of Al$_2$O$_3$ particles
we find $\Delta pK = 4.2$ and $N_S = 0.22/\mathrm{nm}^2$. With these values the 
measured $\zeta$-potential of $52\,$mV at $p\mathrm{H}=6$, $I = 0.01\,$mol/l and 
up to $120\,$mV at $p\mathrm{H}=4$, $I = 0.01\,$mol/l can be reproduced best.
For the experimental determination of the $\zeta$-potential electrophoretic 
(Delsa 440SX, Beckman-Coulter GmbH, Germany) and electrokinetic measurements 
(AcustoSixer IIs, Colloidal Dynamics Ind., USA) were performed. To calculate the 
$\zeta$-potential Henry's theory\cite{Lagaly97} was used. For details see \Ref{Cruz06}.
We have to admit that the relation between the directly measured quantities, e.g.,
electrophoretic mobility, and the $\zeta$-potential is subject of current 
research\cite{Palberg98, Palberg02, Lobaskin04, Lobaskin06}.

\end{section}

\begin{section}{Simulation Setup}
In our simulation we try to model the experimental system as good as 
possible. We start with spherical particles of diameter $d = 0.37\,\mu$m, the mean 
diameter of the particles used in the experiment. The simulation box is 
$48 d = 17.76\,\mu$m long in $x$-direction, $24 d = 8.88\,\mu$m in $z$-direction, and $12 d = 4.44\,\mu$m in $y$-direction. To achieve a volume fraction of usually $\Phi = 35\%$, 
as in the experiment, we need to simulate simulate 9241 spheres.
Our shear direction is the $x$-direction and the velocity gradient of the shear flow 
points in $z$-direction; in other words we shear the upper and lower $xy$-plane with 
respect to each other in $x$-direction. We use periodic boundaries in $x$- and 
$y$-direction and closed boundaries in $z$-direction for both, fluid and MD particles. 
The energy supplied by the shear force is dissipated by means of a Monte-Carlo-thermostat 
described in \Ref{Hecht05} \footnote{Note: The 2 in the denominator of Eq.\,(33) in 
\Ref{Hecht05} hast to be a 3 for the 3D case.}. It acts on the fluid particles as well 
as on the MD particles and conserves the momentum in each SRD cell.
\end{section}

\begin{section}{Experimental Setup}

Experiments are carried out with high purity (99,97\%)
$\alpha$-Al$_2$O$_3$ powder (RCHP DBM, Baikowski Malakoff Industries, Inc., USA).
The mean particle diameter is $0.367\mu\mathrm{m}$ (Coulter LS Particle Size 
Analyzer) and the size distribution is narrow ($d_{10}=0.176\mu\mathrm{m}$, 
$d_{90}=0.664\mu\mathrm{m}$).
The powder is suspended in bidistilled water (Merck, Germany). The suspension is then dispersed with alumina balls in a ceramic container for $24\,$h at a small rotational speed to keep the abrasion low. Subsequently, the suspension is degassed at $50\,$mbar under agitation. Then, in order to reduce the 
ionic strength to the desired degree, the suspension is purified by the dialysis technique. In this way the majority of ions are removed and a background electrolyte of a very low salt concentration ($5\cdot 10^{-4}\,$mol/l) is obtained for suspensions of high solids loading. 
Starting from this master suspension, suspensions with increased ionic strength are obtained by adding different amounts of dry ammonium chloride NH$_4$Cl (Merck, Germany). The $p$H of the suspensions is  adjusted to $p\mathrm{H}=6$ with 0.1 and $1\,$mol/l hydrochloric acid HCl (Merck, Germany), if necessary. 
Thereby the ionic strength and $p$H are revised by use of 
a laboratory $p$H- and conductivity meter (inoLab pH/Cond Level 2, WTW GmbH, Germany).
The electrophoretic mobility of dilute suspensions is measured with a Coulter Delsa 440 SX.
Irreversible aggregation due to inhomogeneous salt concentration is not of importance
here. If the ionic strength is strongly increased and after that, a second dialysis 
step is performed to remove the ions again, the original viscosity is restored.
 
The ion concentrations of selected ions are measured before and after dialysis using Inductively Coupled Plasma-Optical Emission Spectroscopy (ICP-OES, Model JY 70 plus, France). 
The suspensions are characterized using a Viscolab LC10 rheometer 
(Viscolab LM rheometer with control unit Viscolab LC10, Physica, Germany)
with a cup and bob or a double gap geometry. The measurements are 
either performed immediately after suspension preparation, or they are stored on
a roller bank to avoid sedimentation. Sedimentation during the experiment can be
excluded, since it takes much longer than the whole experiment, and the shear forces 
are much larger than gravity.
Shear rate controlled experiments are performed at a constant temperature of 20$^{\circ}$C. The 
suspensions are sheared at a constant shear rate of $\dot\gamma=300/$s ($Pe = 8.8$) before starting 
the actual ramp measurement. In the experiments the shear rate is increased up to
$\dot\gamma=4000\,/$s  ($Pe = 117$) and decreased again to zero. 
In this paper, whenever referring to a shear rate we also specify the P\'eclet number
\begin{equation}
  Pe = 6 \pi \eta R^3 \dot\gamma / k_{\mathrm{B}}T,
\end{equation}
to make it easier for the reader to compare our results to other data.
When the suspensions are pre sheared an occurring discrepancy between the measured viscosity in the increasing ramp and the decreasing one can be minimized. 
A detailed description of the experiments will be published elsewhere\cite{Reinshagen06, Reinshagen06b}.
\end{section}

\begin{section}{Results}
\begin{subsection}{Stability Diagram}

\begin{figure}
\mbox{\epsfig{file=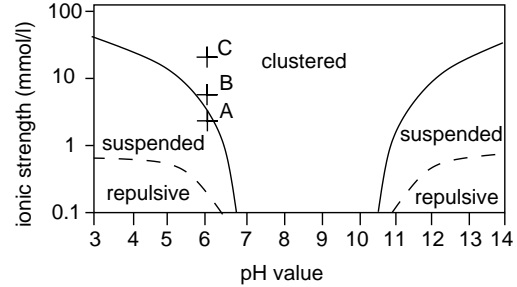,width=\linewidth}}
\caption{Schematic stability diagram for volume fraction $\Phi=35\%$ in
terms of $p$H-value and ionic strength involving three different
microstructures: A clustering regime due to van der Waals attraction, stable  
suspensions where the charge of the colloidal particles prevents
clustering, and a repulsive structure for further increased electrostatic repulsion.
This work concentrates on state $A$ ($p\mathrm{H}=6$, $I=3\,$mmol/l) in the
suspended region, state $B$ ($p\mathrm{H}=6$, $I=7\,$mmol/l) close to the border but 
already in the clustered region, and state $C$ ($p\mathrm{H}=6$, $I=25\,$mmol/l) in the clustered region. The borders are not sharp transitions, but notable in a change of the shear viscosity.
}
\label{fig_stabilitydiag}
\end{figure}

Depending on the experimental conditions, one can obtain three different microstructures: 
A clustered region, a suspended region, and a repulsive structure. The charge regulation model allows us to 
quantitatively relate the interaction potentials to certain experimental conditions. 
A schematic picture of the stability diagram is shown in \fig\ref{fig_stabilitydiag}. 
Close to the isoelectric point ($p\mathrm{H}=8.7$), the particles form clusters for all 
ionic strengths since they are not charged. At lower or higher $p$H values
one can prepare a stable suspension for low ionic strengths because of the charge, which
is carried by the colloidal particles. At even more extreme $p$H values, one can obtain
a repulsive structure due to very strong electrostatic potentials (up to $\zeta = 170\,$mV 
for $p\mathrm{H} = 4$ and $I = 1\,$mmol/l, according to our model). The repulsive structure
is characterized by an increased shear viscosity. In the following we focus on three 
states: State $A$ ($p\mathrm{H}=6$, $I=3\,$mmol/l) is in the suspended region, 
state $B$ ($p\mathrm{H}=6$, $I=7\,$mmol/l) is a point already in the clustered 
region but still close to the border, and state $C$ ($p\mathrm{H}=6$, $I=25\,$mmol/l) is located well in the clustered region.

\begin{figure}
\parbox{0.48\linewidth}{\epsfig{file=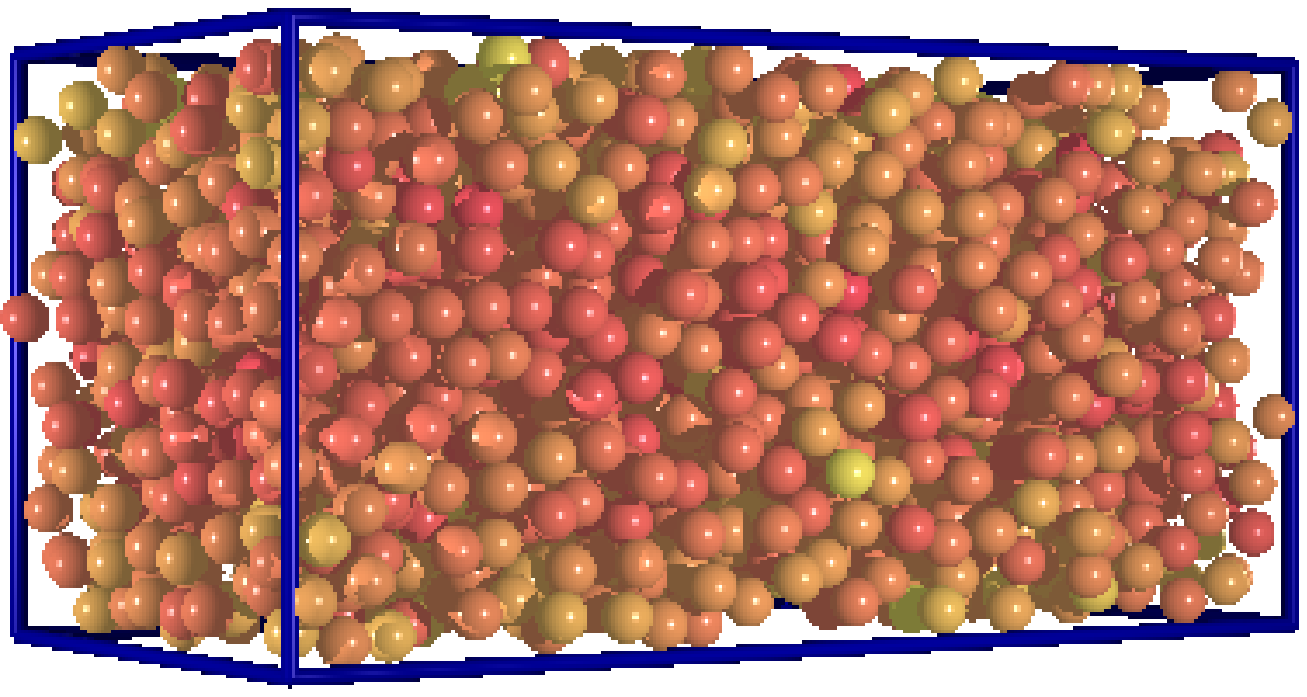,width=\linewidth} \\ a) suspended case}
\parbox{0.48\linewidth}{\epsfig{file=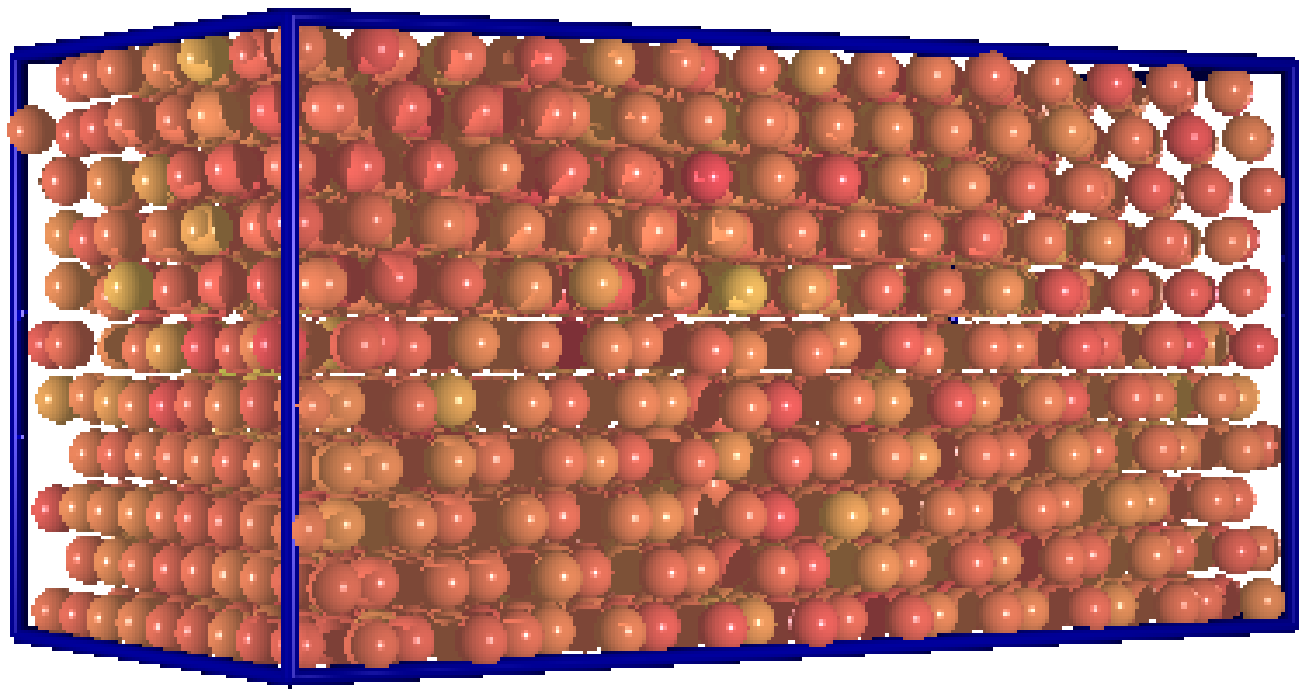,width=\linewidth} \\ b) layer formation}
\parbox{0.48\linewidth}{\epsfig{file=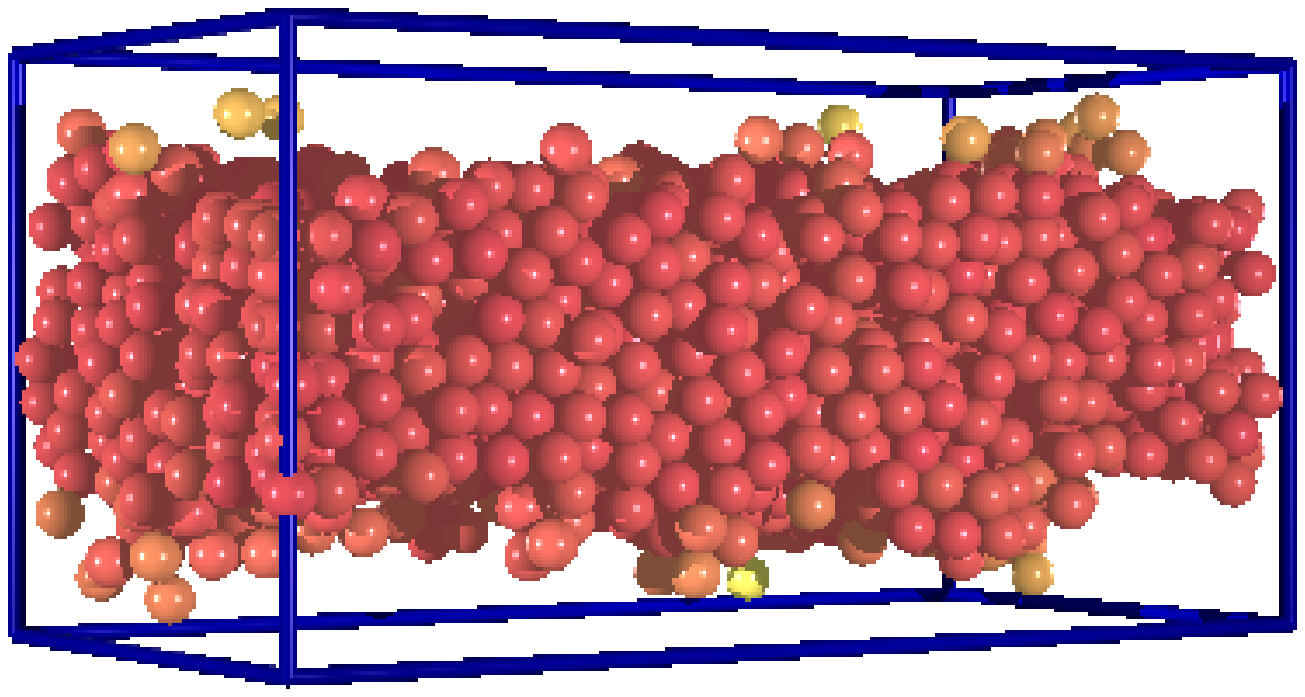,width=\linewidth} \\ c) central cluster}
\parbox{0.48\linewidth}{\epsfig{file=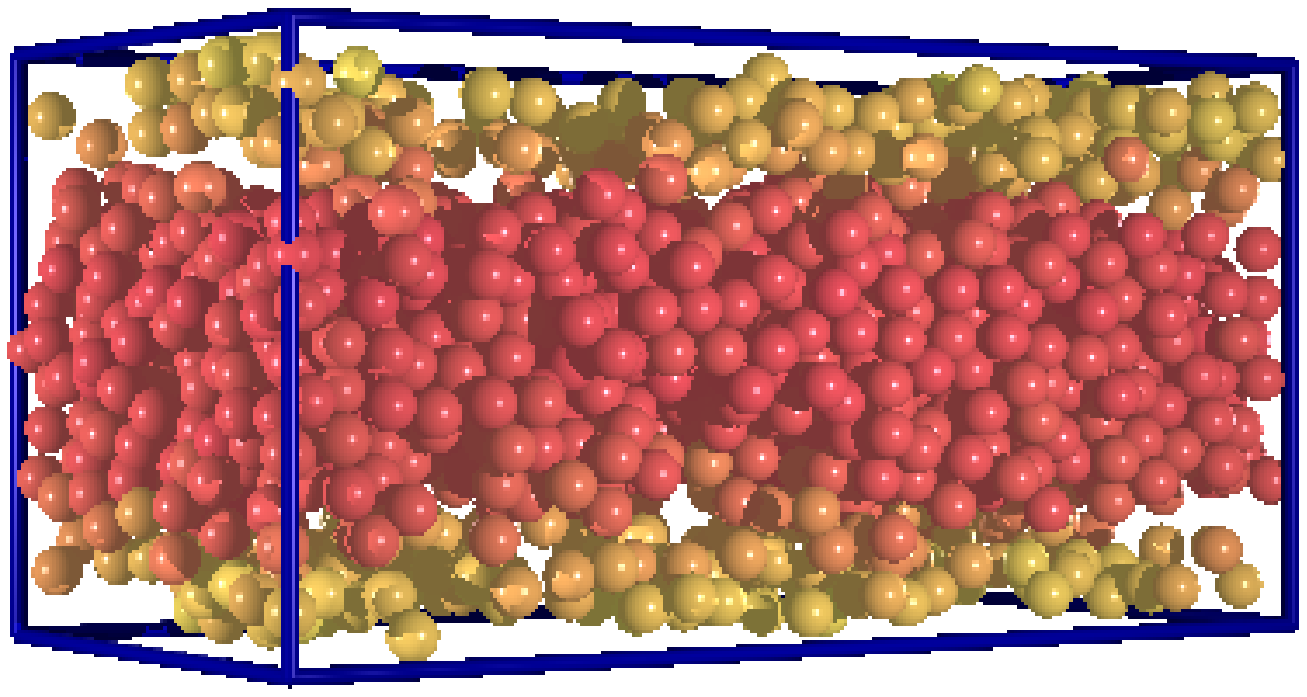,width=\linewidth} \\ d) plug flow}
\caption{(color online) Images of four different cases. For better visibility we have chosen smaller systems than we 
usually use for the calculation of the viscosity. The colors denote velocities: Dark particles are
slow, bright ones move fast. The potentials do not correspond exactly to the
cases $A$--$C$ in \fig\ref{fig_stabilitydiag}, but they show qualitatively the differences between the 
different states: \\
a) Suspension like in state $A$, at low shear rates. \\
b) Layer formation, which occurs in the suspension (state $A$) at high shear rates
and in the repulsive regime already at moderate shear rates.\\
c) Strong clustering, like in state $C$, so that the single cluster in the simulation is formed. \\
d) Weak clustering close to the border like in state $B$, where the cluster can be broken into pieces, which follow the flow of the fluid (plug flow).
}
\label{fig_Snapshots}
\end{figure}

Some typical examples for the different microstructures are shown in 
\fig\ref{fig_Snapshots}a)--d).
These examples are meant to be illustrative only and do not correspond exactly to the
cases $A$--$C$ in \fig\ref{fig_stabilitydiag} denoted by uppercase letters. 
In the suspended case (a), the particles are mainly coupled 
by hydrodynamic interactions. One can find a linear velocity profile and a slight shear thinning. 
If one increases  the shear rate $\dot\gamma>500/$s ($Pe > 15$) , the particles arrange in layers. 
The same can be observed if the Debye-screening length of the electrostatic potential 
is increased (b), which means that the solvent contains less ions ($I < 0.3\,$mmol/l) to screen 
the particle charges. On the other hand, if one increases the salt 
concentration, electrostatic repulsion is screened even more and attractive van der Waals
interaction becomes dominant ($I > 4\,$mmol/l). Then the particles form clusters (c), and 
viscosity rises. A special case, called ``plug flow'', can be observed for high shear rates, 
where it is possible to tear the clusters apart and smaller parts of them follow with the 
flow of the solvent (d). This happens in our simulations for $I = 25\,$mmol/l (state $C$) 
at a shear rate of $\dot\gamma > 500/$s ($Pe > 15$). However, as long as there are only one or two big 
clusters in the system, it is too small to expect quantitative agreement with experiments.
In these cases we have to focus on state $B$ ($I = 7\,$mmol/l) close to the border.

In our simulations we restrict ourselves to the region around $p\mathrm{H}=6$ where we find
the border between the suspended region and the clustered regime at about $I=4\,$mmol/l
in the simulations as well as in the experiments. Also the shear rate dependence of the
viscosity is comparable in simulations and experiments as discussed in \sect\ref{sec_viscosity}.

In \Ref{Graule94} a qualitative stability diagram similar to \fig\ref{fig_stabilitydiag} 
has been shown. The borders there are shifted, since they depend on the threshold value
for which one defines that the viscosity has increased. Correspondingly, if one is 
less sensitive on the viscosity increase, one would still consider the system to be
suspended, if only weak cluster formation takes place.
\end{subsection}


\begin{subsection}{Total Energy}
In our simulations we calculate the total energy, because it can be used as a tool to 
check if the response of the simulation to the variation of any parameter is consistent 
with the expectations, e.g., a decrease of the surface charge on the colloidal particles
should cause the secondary minimum of the DLVO potential to become deeper and thus decrease 
the total energy--but if the total energy increases, this can be an indication for numerical
instabilities.

The total energy comprises the kinetic energy of both fluid and colloidal particles, 
including thermal motion on the microscopic level, as well as the potential energy due 
to Coulomb repulsion, van der Waals attraction, and Hertz contact forces. Our simulations 
are carried out at room temperature ($T = 295\,$K) and constant volume fraction. 
Supposing a linear velocity profile, the kinetic energy increases quadratically with 
the shear rate $\dot\gamma$. This can be observed if the electrostatic repulsion is, 
on the one hand, strong enough to prevent cluster formation due to van der Waals attraction 
and, on the other hand, weak enough, so that the colloidal particles can move 
relatively freely without undergoing a glass transition or crystallization.

If the interactions are strongly repulsive, i.e., in the case of very low salt concentration,
where the Debye-screening length is large, one can see an extra contribution of the 
electrostatic repulsion to the total energy. If the volume fraction is low, the particles
can still find a configuration, in which the mean nearest neighbor distance is larger than 
the interaction range of the repulsion. But, if the volume fraction is increased, 
the particles have to be packed closer, which leads to a constant positive offset to the total 
energy. It only depends on the potentials and on the volume fraction, but not on the shear rate.

\begin{figure}
\mbox{\epsfig{file=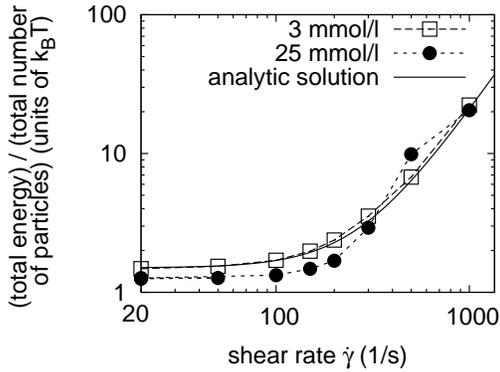,width=\linewidth}}
\caption{Total energy depending on the shear rate $\dot\gamma$ for the states $A$ ($I=3\,$mmol/l)
and $C$ ($I=25\,$mmol/l) of \fig\ref{fig_stabilitydiag}. In state $A$ the system is a stable suspension, 
in state $C$ cluster formation reduces the total energy at low shear rates. At $\dot\gamma = 500/$s ($Pe = 15$),
the cluster can be broken up into two parts moving in opposite directions. The two solid bodies have
a larger kinetic energy than the suspension with a linear velocity profile. This explains the crossover of the two curves. For even higher $\dot\gamma$, the clusters are broken up in more pieces leading 
ultimately to the same structure as for the suspended state $A$. The energy axis has been scaled 
by the total number of particles (fluid particles plus colloidal particles) and plotted in units
of $k_{\mathrm{B}}T$. The solid line is the analytical solution (\eq\ref{eq_analytic}) for a linear 
velocity profile, the dashed lines are a guide to the eye.
}
\label{fig_energy}
\end{figure}

In a similar way as for repulsive interactions, one can understand the negative energy contribution 
in the case of high salt concentrations: The DLVO potentials contain a minimum where attractive 
Van der Waals interaction is stronger than electrostatic repulsion. Then the particles form 
clusters and ``try'' to  minimize their energy. In \fig\ref{fig_energy} for small shear rates
the values for the energy in the clustered case of state $C$ ($I = 25\,$mmol/l) is lower than for 
the suspended case of state $A$ ($I = 3\,$mmol/l). We have plotted the total energy divided 
by the total number of all particles (fluid particles plus colloidal particles) in units of 
$k_{\mathrm{B}}T$. For $\dot\gamma \rightarrow 0$ the energy per particle approaches $\frac{3}{2}k_{\mathrm{B}}T$ as one would expect in 3D. The solid line in \fig\ref{fig_energy} 
is the sum of $\frac{3}{2}k_{\mathrm{B}}T$ per particle with the kinetic energy of a fluid with a 
linear velocity profile:
\begin{equation}
\label{eq_analytic}
 E_{\mathrm{tot}} = \frac{3}{2}k_{\mathrm{B}}T N_{\mathrm{tot}}+
   \frac{1}{24}V\bar\rho \dot\gamma^2 L_z^2,  
\end{equation}
where $N_{\mathrm{tot}}$ is the total number of both, fluid and colloidal particles, $V$ 
denotes the volume of the simulated system, $\bar\rho$ the averaged mass density of the suspension, 
and $L_z$ is the extension in $z$-direction (perpendicular to the shear plane).
State $A$ coincides very well with this curve.

For state $C$ the behavior is shear rate dependent: In contrast to the repulsive case, 
clusters can be broken up. This happens at
a shear rate $\dot\gamma=500/$s ($Pe = 15$) (see \fig\ref{fig_energy}) where one obtains two clusters moving in 
opposite directions. Since in this case the resistance of the system decreases, the velocity of 
the two clusters becomes larger. Since both clusters are moved as a whole, their energy even 
becomes larger than in the suspended case. 
If one further increases the shear rate, no (big) clusters can form 
anymore and the energies for both salt concentrations are nearly the same, and correspond to 
the kinetic energy of a suspension with a nearly linear velocity profile. For $\dot\gamma=1000/$s ($Pe = 29$)
state $C$ coincides with the analytic curve.
\end{subsection}

\begin{subsection}{Shear Profile and Shear Viscosity}
\label{sec_viscosity}
\begin{figure}
\epsfig{file=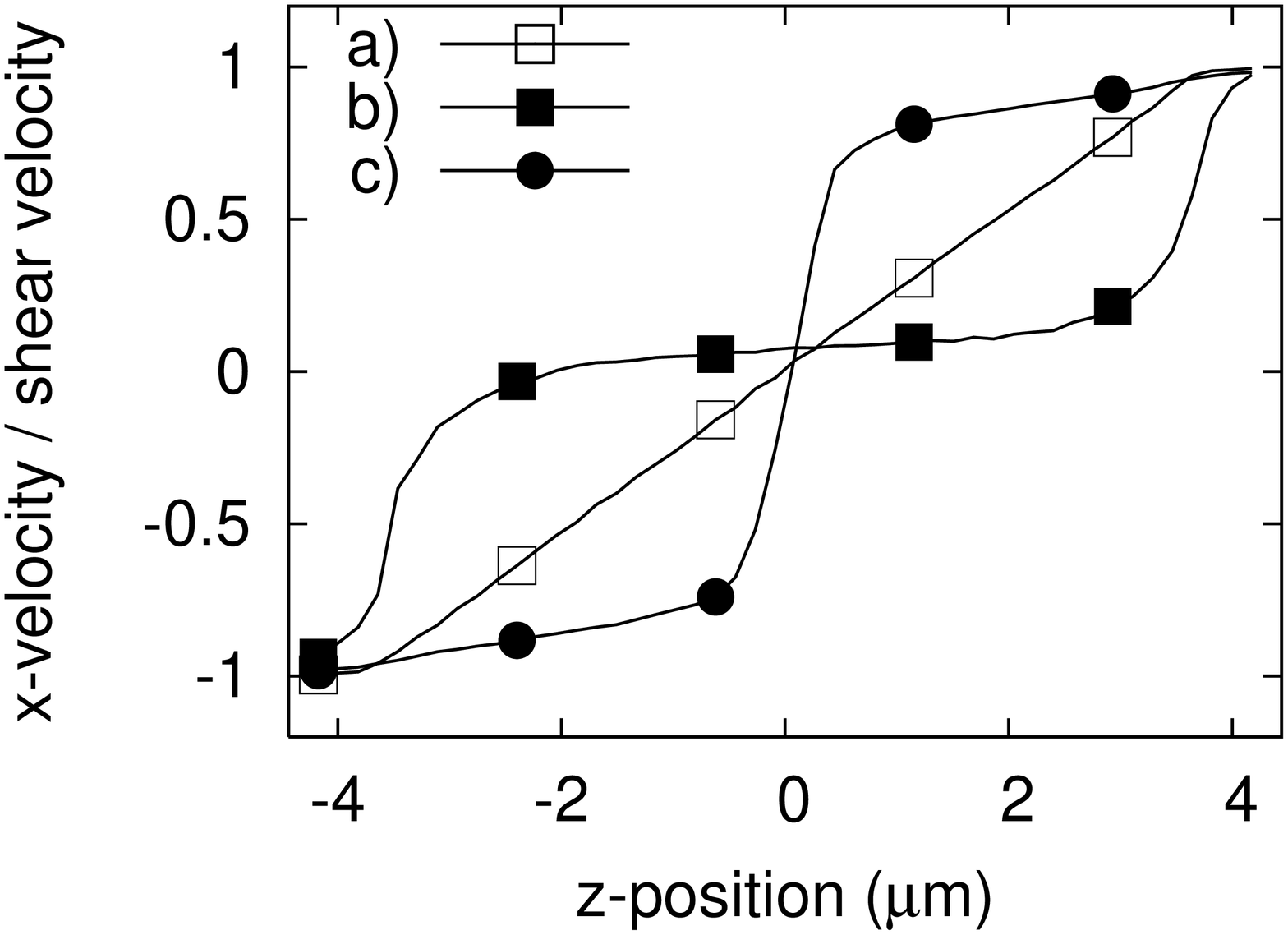,width=\linewidth}
\caption{Profiles of tangential velocity component ($v_x$) in normal direction ($z$): \\
a) Linear profile in the suspended regime, state $A$ of \fig\ref{fig_stabilitydiag} ($I=3\,$mmol/l) at  $\dot\gamma=500/$s ($Pe = 15$)). \\
b) Cluster formation in state $C$ ($I=25\,$mmol/l) at $\dot\gamma=100/$s ($Pe = 2.9$). 
In principle one could determine the viscosity of one single cluster from the central plateau, 
but this is not the viscosity found in experiments. There, one measures the viscosity of a paste
consisting of many of these clusters. \\
c) Same as case b) but with higher shear rate ($\dot\gamma=500/$s $Pe = 15$). 
Hydrodynamic forces are large enough to break the cluster into two pieces. \\
The velocity axis is scaled with the shear velocity $v_S$ for better comparability.}
\label{fig_VxProfiles}
\end{figure}
\begin{figure*}
\mbox{a)\epsfig{file=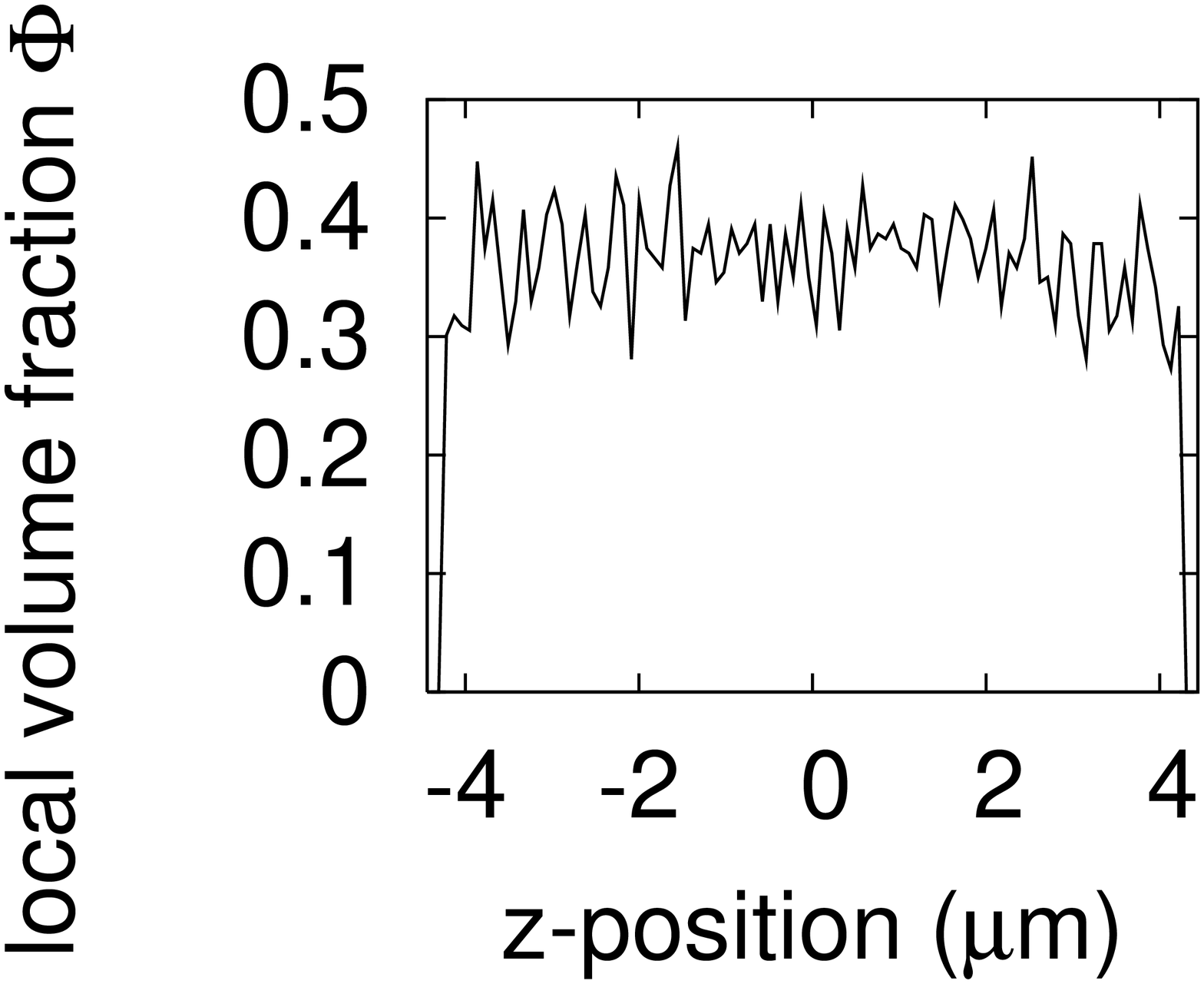,width=0.3\linewidth}
b)\epsfig{file=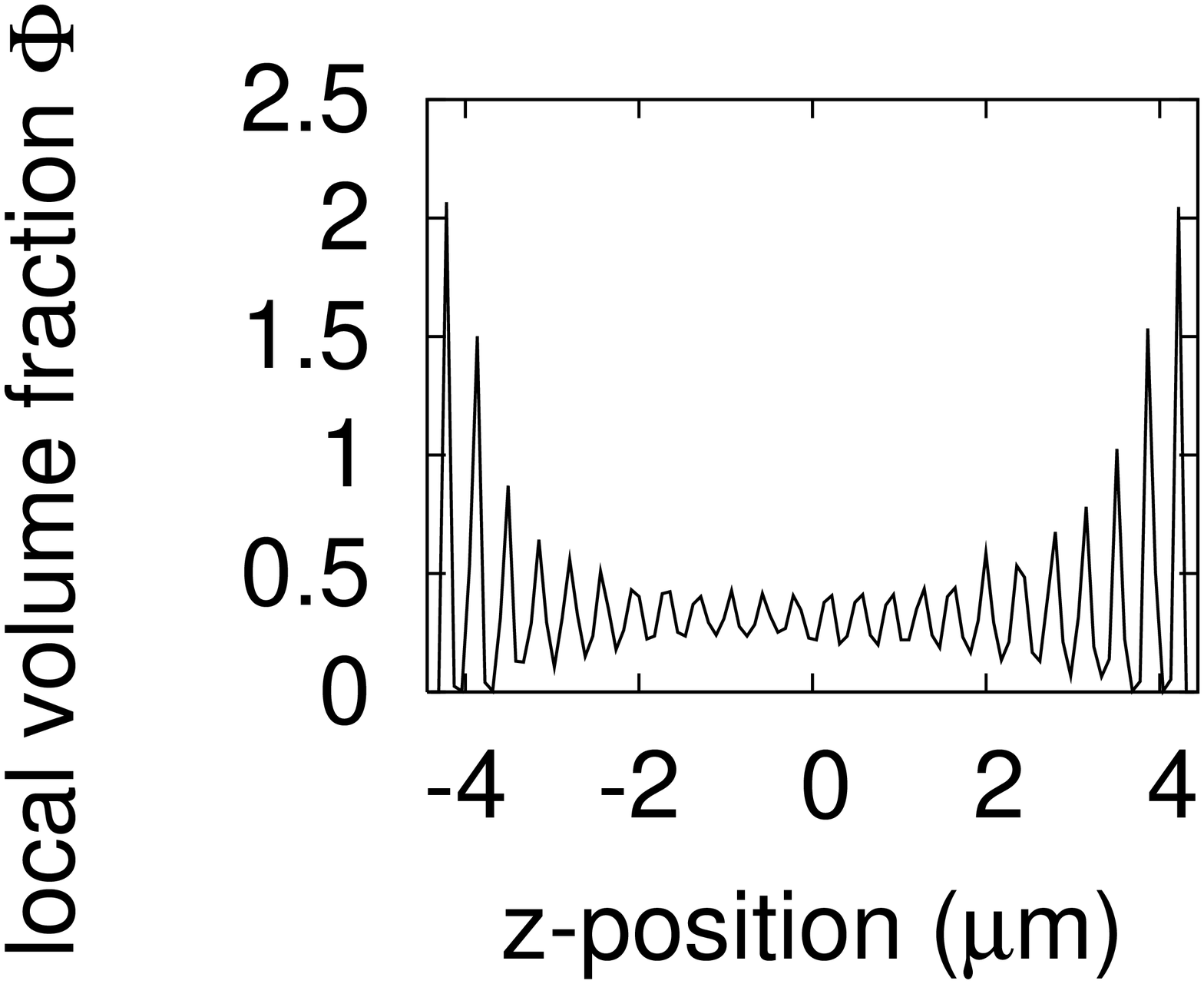,width=0.3\linewidth}
c)\epsfig{file=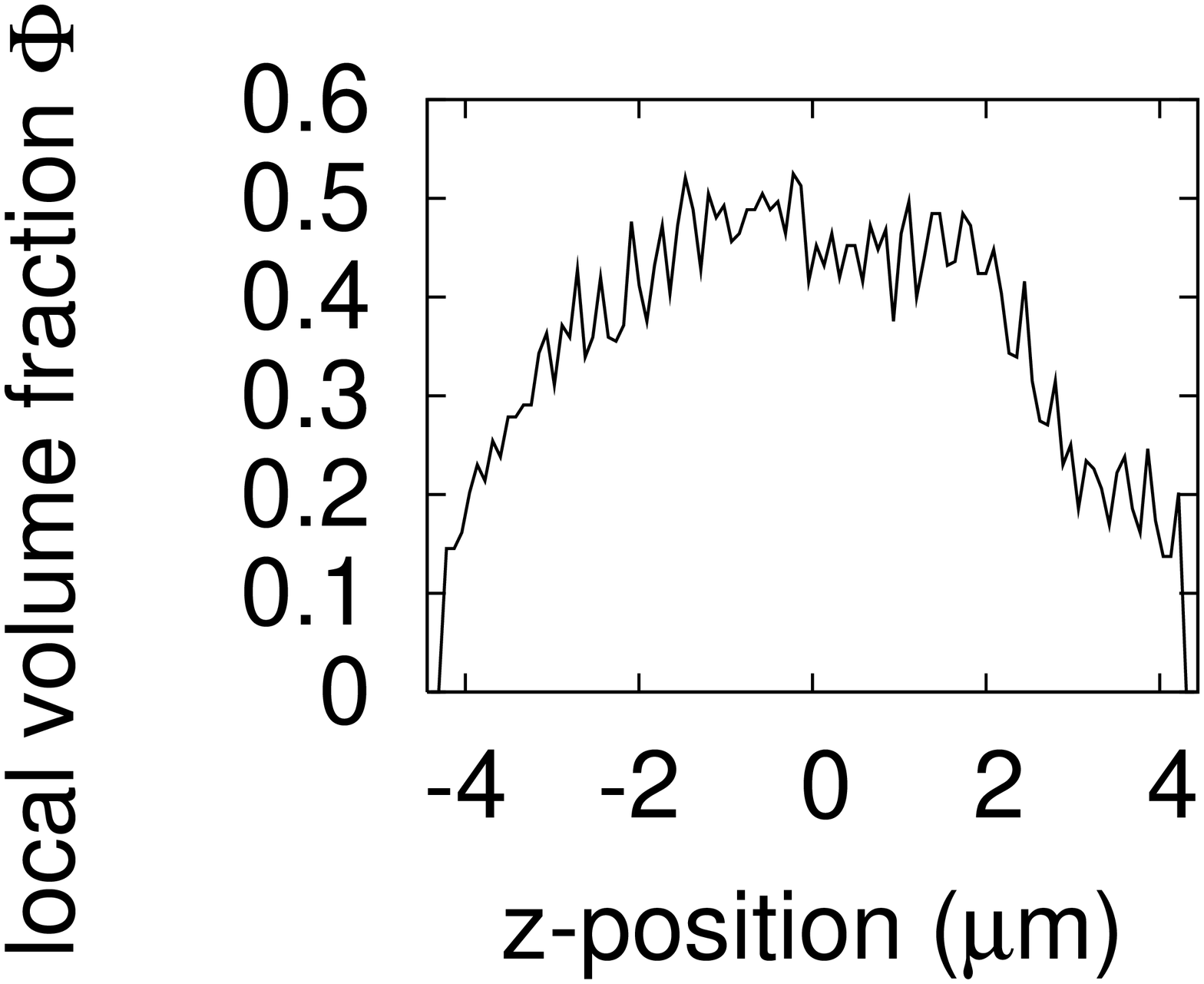,width=0.3\linewidth}
}
\caption{Density profiles: a) Suspended case: State $A$ in \fig\ref{fig_stabilitydiag} ($I=3\,$mmol/l), at low shear rates ($\dot\gamma=50/$s ($Pe = 1.5$)). The density distribution is homogeneous. \\
b) Shear induced layer formation: This is state $A$ as in graph a) of this figure, but for a high shear rate ($\dot\gamma=1000/$s ($Pe = 29$)). \\
c) Strong attractive forces in state $C$ ($I=25\,$mmol/l): For low shear rates ($\dot\gamma=50/$s ($Pe = 1.5$)) only one central cluster is formed, which is deformed slowly.
}
\label{fig_densityProfiles}
\end{figure*}
\begin{figure}
\mbox{\epsfig{file=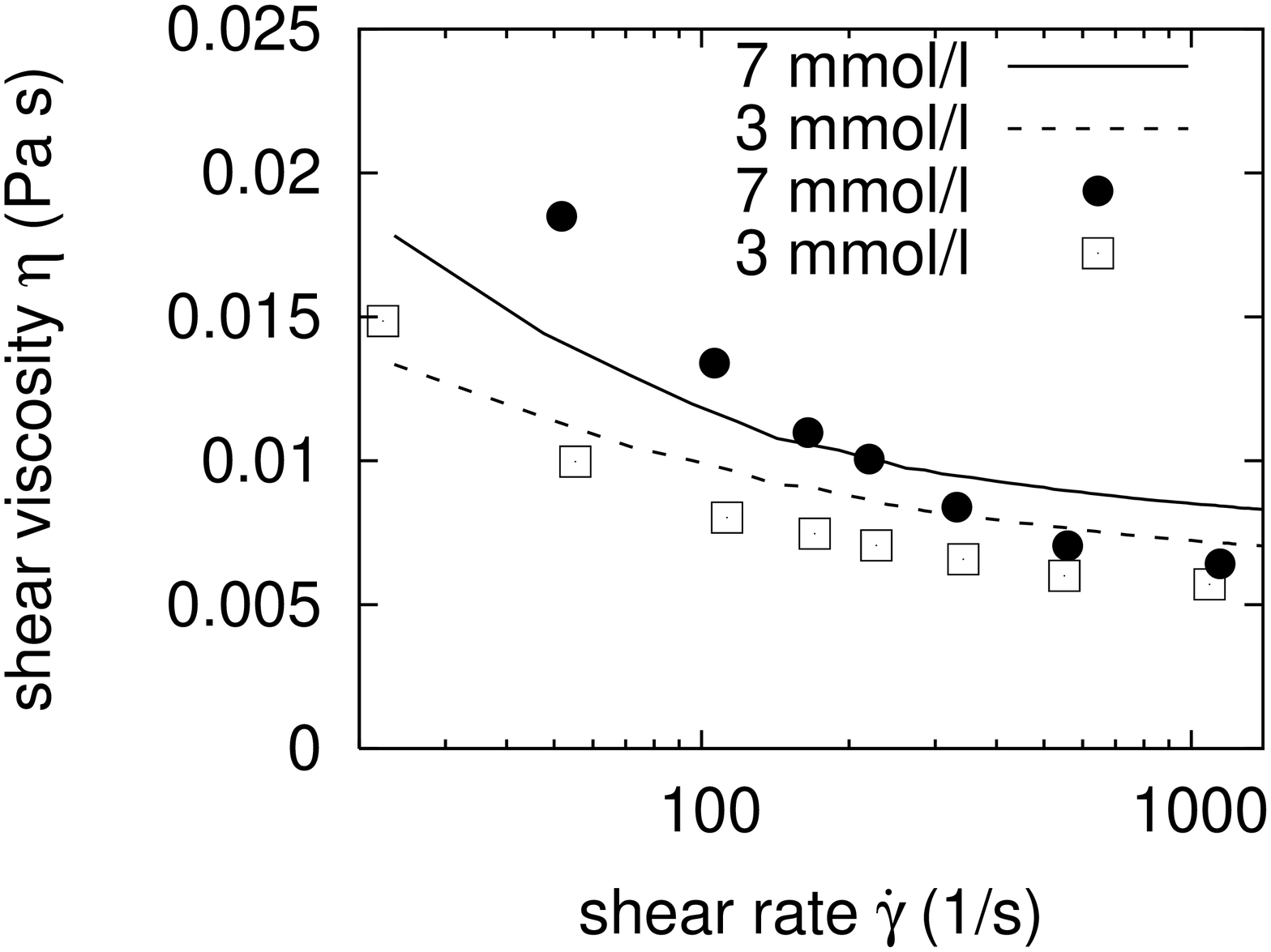,width=\linewidth}}
\caption{Comparison between simulation and experiment: viscosity in dependence of the shear rate. 
for the states $A$ ($I=3\,$mmol/l) and $B$ ($I=7\,$mmol/l) of \fig\ref{fig_stabilitydiag}. 
Note: shear thinning is more pronounced for the slightly attractive 
interactions in state $B$ than for the suspended state $A$. Lines denote experimental data 
\cite{Reinshagen06}, points are results from our simulations.}
\label{fig_viscosity}
\end{figure}
In each of the three regimes a typical velocity profile of the shear flow occurs. 
For the suspended microstructure one finds a linear velocity profile 
(\fig\ref{fig_VxProfiles}a)) with nearly Newtonian flow. The particles are distributed 
homogeneously, thus the density profile is structureless (\fig\ref{fig_densityProfiles}a)).
The motion of the particles is only weakly coupled by the hydrodynamic forces. 
At high enough shear rates ($\dot\gamma > 500/$s ($Pe > 15$)) the  particles arrange in 
layers parallel to the shear plane, as can be seen in the density profile
 \fig\ref{fig_densityProfiles}b), too.
This arrangement minimizes collisions between the particles. As a result, the shear viscosity 
descents as shown in \fig\ref{fig_viscosity}, which we discuss more in detail below. 
Shear induced layer formation has been reported in literature for different 
experiments\cite{Ackerson85, Ackerson88, Ackerson89, Cohen04, Vermant05} and 
Stokesian dynamics simulations\cite{Brady96, Brady00}. For low shear rates Brownian motion
disturbs the layers or prevents their formation. As shown in \Ref{Brady96}, hydrodynamic
forces can destroy them as well, if the shear rate is high enough. In our simulations
we do not reach these conditions. In the simulations shear rates up to $2000/$s ($Pe = 59$) 
can be realized before limitations of the simulation method influence the results. The 
increment of the shear angle in one SRD time step $\gamma = \dot\gamma \tau_{SRD}$ 
amounts about $\pi/4$ then, i.e., the offset in $x$-direction between two neighboring 
layers of SRD cells in $z$-direction amounts one cell per SRD time step.

Furthermore, the volume fraction and the interaction
range of the electrostatic repulsion, or the ionic strength respectively, influence the
layer formation: In the repulsive regime the layers are more clearly already at moderate 
shear rates. It can be excluded that the effect is purely a finite size effect, since
for un-sheared suspensions no layers can be observed, at least some particle diameters from 
the walls. In the repulsive regime the particles try to optimize their local structure, 
but a long range order as in the case of a sheared system can not be seen. 

In the clustered regime, the clusters move in the fluid as a whole. They are deformed, but
since the inter-particle forces are stronger than the hydrodynamic forces, the 
cluster moves more like a solid body than like a fluid. 
Often there is one big cluster that spans the whole system. The density profile 
(\fig\ref{fig_densityProfiles}c)) increases in the central region and decays at the 
regions close to the border, since particles from there join the central cluster.
When averaging the velocity profile in the shear flow, one finds a very small velocity 
gradient in the center of the shear cell and fast moving particles close to the wall, 
where the shear is imposed (\fig\ref{fig_VxProfiles}b)). The velocity profile is 
non-linear on the length scale of the simulations. 
In the experiment the physical dimensions are much larger and therefore the velocity profile
can become approximately linear again if the system consists of many large clusters.
However, due to the computational effort in simulations 
it is today impossible to measure the shear viscosity for these strongly inhomogeneous systems.
We have scaled
our system by a factor of 2 in $x$ and $z$-direction (keeping the volume fraction $\Phi = 35\%$ 
constant), but we still observe one big cluster after some hundreds of SRD time steps,
i.e., finite size effects are still present in our simulations.

Closer to the border clusters can then be broken up into small pieces by the 
hydrodynamic forces at least for high shear rates. In state $C$ of \fig\ref{fig_stabilitydiag}
this happens for the first time at $\dot\gamma = 500/$s ($Pe = 15$), so that one can find two clusters in 
the system moving in opposite directions. The velocity profile of this case is shown in
\fig\ref{fig_VxProfiles}c). For even higher shear rates or closer to the  
border (e.g. state $B$), the clusters are broken into smaller pieces. Then, they 
move in the shear flow with an approximately linear velocity profile. 
Due to van der Waals attraction the system resists with stronger shear forces and the 
viscosity is higher than in the suspended case (\fig\ref{fig_viscosity}).

In \fig\ref{fig_viscosity} the simulation results are shown together with the experimental
results, both for the two cases of a slightly clustered system in state $B$ ($I = 7\,$mmol/l) 
and a suspension (state $A$, $I = 3\,$mmol/l). 
For the suspension (state $A$) the viscosity decreases with the shear 
rate (``shear thinning''). The experimental data and the simulation are 
consistent within the accuracy of our model. There are several 
reasons for which our model does not fit exactly the measurements:
The most insecure factor which enters into the comparison is the measurement of 
the $\zeta$-potential. Starting from this point we set up our charge regulation model
to extrapolate to different salt concentrations, assuming two reactions being the 
only processes that determine the surface charge of the colloidal particles. Furthermore, 
we have monodisperse spherical particles, which is another simplification in our model. 
Then, the lubrication force as a correction for the finite resolution of the fluid method 
can only recover to a certain degree the hydrodynamics on smaller length scales than the 
cell size of the fluid simulation, e.g., we have not implemented other modes of 
lubrication than the ``squeezing mode'' (\eq\ref{eq_FLub}). 

However, we have done several tests where we have simulated systems with 
size and charge polydispersity in the order of magnitude corresponding to the 
experimental conditions. We have tested different sorts of boundary conditions, different
ways to implement shear, and different coupling methods between fluid and particles.
In most of the tests the achieved shear viscosity in the simulation did not change notably.
To our experience the way shear is imposed and the particle size have the largest influence 
on the result.

Finally, one has to keep in
mind that the viscosity of the suspension can be varied by more than one order of
magnitude, e.g., by changing the ionic strength. In this context the deviations between 
simulation and experiment are small.

\begin{figure}
\mbox{\epsfig{file=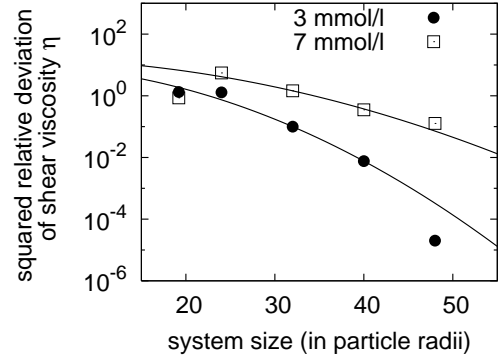,width=\linewidth}}
\caption{Discrepancy between simulated viscosity and measurement for states $A$ and $B$ of \fig\ref{fig_stabilitydiag} for different system sizes at low shear rates. The plot shows 
squared relative differences against $z$-extension of the simulation volume. The lines 
are a guide to the eye.}
\label{fig_finite}
\end{figure}

For the slightly clustered case (state $B$) an increase of the shear viscosity, 
compared to the suspended case, can be observed in the experiment as well as in the simulations. 
Shear thinning becomes more pronounced, because clusters are broken up,
as mentioned above. However, the shear rate dependence is stronger in the simulations
than in the experiment. This can be the first indication of finite size effects. 

We have studied the dependence of the simulated shear viscosity in dependence of the system 
size. The effect is most important for low shear rates and thus we carried out several 
simulations for state $A$ at $\dot\gamma = 20/$s ($Pe = 0.6$) and for state $B$ at 
$\dot\gamma = 50 /$s ($Pe = 1.4$) We have chosen these values because clustering is already too 
strong in state $B$ at $\dot\gamma = 20/$s ($Pe = 0.6$) to reasonably determine a viscosity 
and the system size dependence becomes too small for  $\dot\gamma = 50/$s ($Pe = 1.4$) in state $A$.
In \fig\ref{fig_finite} we plot the squared relative deviation between simulation and 
measurements against the system size. We do not know, if the simulation results would 
exactly converge to the measured values if the simulated system is large enough.
However, the figure shows the trend that the deviation becomes smaller for larger system sizes,
but to reach in state $B$ the same accuracy as in state $A$ one would have at least to double 
the system size in each dimension. It then takes approximately twice as long for the system
to relax to a steady state, resulting in a factor of 16 in the computational effort. This 
gives a guess, that each single point of \fig\ref{fig_viscosity} would need approximately 3000 CPU 
hours. For smaller shear rates or even deeper in the clustered regime of the stability diagram, 
e.g., in state $C$ ($I = 25\,$mmol/l), the  finite size effects become more pronounced---ending up 
in the extreme case of only one big cluster existing in the system. For simulations with good 
accuracy the effort again increases at least by the same factor.

Unfortunately, exactly this would be the most interesting case with respect to soil mechanics
and to understand landslides, which was our initial motivation. Anyhow, if we compare state $A$
and state $B$, shear thinning becomes stronger with increasing ionic strength. In the 
experiments the effect becomes much stronger for larger ionic strengths (up to $I = 65\,$mmol/l),
where the viscosity for low shear rates is increased by more than a factor of ten. However,
the fact that there is shear thinning and that it depends on the ionic strength is an 
interesting result, if we recall the fact that the lubrication force in our simulation 
can be interpreted as a velocity dependent damping force, which becomes {\emph stronger} for
higher relative velocities. Therefore one would intuitively expect shear thickening 

Finally, the limitations of the DLVO theory have to be taken into account. DLVO potentials are 
derived for dilute suspensions and hence large particle distances. This is not fulfilled in 
our case--and even less inside the clusters. There are theoretical attempts that address the 
shortcomings of DLVO theory: Explicit simulation of micro-ions\cite{Linse00}, 
density functional theory\cite{Loewen98, vanRoij97, vanRoij99}, 
response theory\cite{Grimson91, Denton99, Denton00, Denton04},
Poisson Boltzmann cell models\cite{vonGruenberg01, Deserno02, Tamashiro03},
and full Poisson Boltzmann theory\cite{Bowen98, Gray99, Dobnikar03, Dobnikar03b}, 
but they have other disadvantages--most of them require a large computational effort. 
For Poisson Boltzmann cell models one assumes homogeneously distributed colloidal particles,
so that each of them can be regarded as a representative single particle in a Wigner-Seitz cell. 
Additionally, depending on the level Poisson Boltzmann theory is included, a mysterious phase 
separation could be identified as an artifact of linearization\cite{Tamashiro03}.
Full Poisson Boltzmann theory would require the calculation of the local potential, 
not only as done in our charge regulation model in the beginning of the simulation, but 
for the whole simulation box and in each time step. This would in principle provide a 
better description of the real system, but the computational effort would be much larger
making simulations of several thousands of particles impossible. The same applies to the 
approach of including the micro-ions explicitly in the simulation. One could obtain 
three (many) body interactions from full Poisson Boltzmann theory and try to include them as 
a lookup table in the simulation. However, one would have to decrease the system size 
to keep the computational effort affordable.
For our simulations we need relatively simple pair potentials to keep the computational costs
within a limit. Nevertheless, the overall behavior can be reproduced by the simulation on a 
semiquantitative level. The reason for that might be the fact that in some of the above mentioned 
theoretical attempts (density functional theory and cell models) DLVO-like potentials are obtained
with a renormalized charge and screening length. In our charge regulation model we do
nothing else than adjusting a renormalized charge to the measurements of the $\zeta$-potential.
This may be a general explanation why DLVO potentials can often be used although the
assumptions for DLVO theory are not fulfilled\cite{Alexander84, Palberg95, Kjellander92, Trizac02c, Trizac03}.

We have carried out simulations in the repulsive region of the stability diagram as well. 
We find layers parallel to the shear plane in analogy to \fig\ref{fig_densityProfiles}b). 
In contrast to the suspended regime, in the repulsive regime the layer structure is present---at 
least locally, but orientationally disordered---even if no shear is applied. If shear flow is 
present, the shear plane marks one 
orientation which the layer structure adopts. In some cases for very low ionic strengths one can 
observe shear bands so that the velocity gradient and thus the viscosity as well vary strongly in the system. Again, in the experiment,  physical dimensions are much larger and on that length scale the 
velocity profile might be assumed to be linear when enough shear bands are in the system.
The shear force and hence the viscosity rise with respect to the suspended regime, due to 
electrostatic repulsion. One can consider the particles together with the interaction range as 
soft spheres with an effective radius of the interaction range of the electrostatic repulsion. 
This  effective radius can be in our case about $25\%$ larger than the particle radius. 
Therefore, a transition to a repulsive structure occurs in our systems already 
between $35\%$ and $40\%$ volume fraction. Because of the smooth shape of the exponentially 
screened Coulomb  potential it is not a sharp glass transition as for hard spheres, but 
smooth and shear rate dependent as well. In \fig\ref{fig_glassy} we have shown the 
dependence of the viscosity on the volume fraction for $p\mathrm{H} = 6$ and $I = 0.3\,$mmol/l.
Starting at $\Phi = 0.3$ the shear viscosity starts to increase and reaches a value one 
decade larger beyond $\Phi = 0.4$.
\begin{figure}
\mbox{\epsfig{file=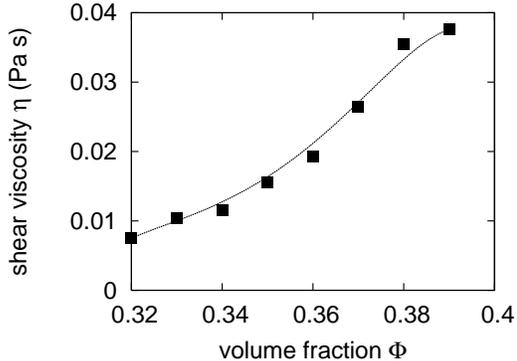,width=\linewidth}}
\caption{Viscosity versus volume fraction for the repulsive region ($p\mathrm{H} = 6$ and 
$I = 0.3\,$mmol/l). The shear rate was $\dot\gamma = 100/$s ($Pe = 2.9$). 
The points are simulation results, the line is a guide to the eye.}
\label{fig_glassy}
\end{figure}
\end{subsection}
\end{section}

\begin{section}{Summary and outlook}
We have shown how to relate DLVO potentials to the conditions in a real 
aqueous suspension of Al$_2$O$_3$ particles. The behavior of shear viscosity
has been studied in experiments and in simulations. We have found shear 
thinning due to a layer formation on the microscopic scale in the case of a 
suspension.

If a clustered system is sheared, clusters are broken up into pieces by the 
imposed shear, which leads to a stronger shear thinning than in the suspended 
case. Close to the border we are able to reproduce the measured 
shear viscosity in the simulation. 

Deep in the clustered regime we have found that our particles form one big
cluster in the system which can be broken up by the hydrodynamic forces of 
the shear flow. For strongly clustered systems at low shear rate,
which would be the most interesting case for soil mechanics, there
are strong finite size effects. One attempt to address this problem
is to increase the size of the simulated system. As we have shown with 
our work, the computational effort increases to an extend that 
a parallelized simulation code would be necessary. However, 
this might be not sufficient, since in both cases, for low shear rates and for 
strongly attractive interactions the finite size effects become stronger. 
Since in these simulations a considerable ammount of the computing time 
is consumed by the particles inside a cluster, one could think of a 
more coarse-grained description of the clusters. Nevertheless, input data 
for such a model could be obtained using our present simulation code. 
Depending on the model one might need information about the shear resistance
of a single cluster, depending on the cluster size and shape. This would be
very difficult to measure, but it could be calculated in a small simulation
using our combined MD and SRD algorithm.
\end{section}

\begin{acknowledgments}
This work has been financed by the German Research Foundation (DFG) within the project
DFG-FOR 371 ``Peloide''. We thank G.~Gudehus, G.~Huber, M.~K\"ulzer, L.~Harnau, 
and S.~Richter for valuable collaboration. 

Parts of this work are resulting from the collaboration with the group of 
A.~Coniglio, Naples, Italy. M.~Hecht thanks him and his group for his hospitality and 
for the valuable support during his stay there. M.~Hecht thanks the DAAD for the 
scholarship (Doktorandenstipendium) which enabled him the stay.

H.~J.~Herrmann thanks the Max Planck prize.

M.~Bier is deeply indebted to M.~Biesheuvel and S.-H.~Behrens for helpful
discussions on charge regulation models. 

The computations were performed on the IBM p690 cluster at the Forschungszentrum 
J\"{u}lich, Germany. 

\end{acknowledgments}


\end{document}